# UrbanVCA: a vector-based cellular automata framework to simulate the urban land-use change at the land-parcel level

**Authors:** Yao Yao[a], Linlong Li[a], Zhaotang Liang[b], Tao Cheng[a], Zhenhui Sun[a], Peng Luo[c], Qingfeng Guan[a,*], Yaqian Zhai[a], Shihao Kou[a], Yuyang Cai[a], Lefei Li[d], Xinyue Ye[e,*]

[a] School of Geography and Information Engineering, China University of Geosciences, Wuhan, China.
[b] Institute of Space and Earth Information Science, The Chinese University of Hong Kong, Hong Kong.
[c] Chair of Cartography, Technical University of Munich, Munich, Germany.
[d] Didi Chuxing Technology Co., Beijing, China.
[e] Landscape Architecture and Urban Planning, Texas A&M University, College Station, TX, United State.

[*] Corresponding Authors: Dr. Qingfeng Guan (guanqf@cug.edu.cn) and Dr. Xinyue Ye (xinyue.ye@tamu.edu)

*Abstract*: Vector-based cellular automata (CA) based on real land-parcel has become an important trend in current urban development simulation studies. Compared with raster-based and parcel-based CA models, vector CA models are difficult to be widely used because of their complex data structures and technical difficulties. The UrbanVCA, a brand-new vector CA-based urban development simulation framework was proposed in this study, which supports multiple machine-learning models. To measure the simulation accuracy better, this study also first proposes a vector-based landscape index (VecLI) model based on the real land-parcels. Using Shunde, Guangdong as the study area, the UrbanVCA simulates multiple types of urban land-use changes at the land-parcel level have achieved a high accuracy (FoM=0.243) and the landscape index similarity reaches 87.3%. The simulation results in 2030 show that the eco-protection scenario can promote urban agglomeration and reduce ecological aggression and loss of arable land by at least 60%. Besides, we have developed and released UrbanVCA software for urban planners and researchers.

*Highlights*:
1. An UrbanVCA software supporting land-use change simulation at land-parcel level is designed.
2. This study firstly proposed and implemented a vector-based landscape index.
3. UrbanVCA can effectively simulate urban expansion and internal land-use changes.
4. UrbanVCA can obtain high simulation accuracy of urban land-use change.
5. The eco-protection scenario can reduce ecological aggression and loss of arable land by at least 60% in 2030.

# 1. Introduction

China has experienced rapid urbanization in the past several decades (Chen 2007; Liu et al. 2010). The urban development mode of China has gradually changed from outward urban expansion to internal urban renewal(Jiang et al. 2017). The renewal of urban space is an internal land-use change at a land parcel scale (Barreira-González, Gómez-Delgado, and Aguilera-Benavente 2015). How to accurately simulate the process of urban development, especially for urban renewal, has become a hot topic in urban development simulation studies.

Cellular automata (CA) based simulation model is the most popular method for fine-scale urban simulation at present, which can be used to understand the process of urban development and predict the future trend (Clarke, Hoppen, and Gaydos 1997; Santé et al. 2010; Dahal and Chow 2015; Feng and Tong 2020). The traditional CA model uses the same-size grid unit as a cell lattice, like raster pixels or patches, to simulate the spatiotemporal dynamics of land-use structure (Liu and Phinn 2003; Chen et al. 2014; Tian et al. 2016; Kang et al. 2019). The state-of-art development of such type of CA model was mixed-cell CA (Liang et al. 2021). Within a grid cell, mixed-cell CA was composed of an array of continuously valued land-use type and is able to simulate the continuous change of multiple land-use components at a sub-cell scale. While mixed-CA was able to simulate the subtle changes of land-use proportions within a land unit, its regular cell lattices remain simplified to differentiate geographic entities and consider the morphological change.

Physical settings of geographic entities are normally irregular polygons and disagree with simplified grid-cell lattice settings in conventional CAs (Rabbani, Aghababaee, and Rajabi 2012). For the finer granularity of CA simulation, grid cell lattice fails to accommodate complicated land-use transitions with highly irregular, fragmented, and inconsistent geometric entities in urban areas (Moreno, Wang, and Marceau 2010). Therefore, irregular cell-based vector CAs (VCA) has been proposed and applied to urban land-use change simulations, including cell lattices of Tyson polygon, Delaunay triangulation, street plot, and cadastral plot (Semboloni 2000; Moreno, Ménard, and Marceau 2008; Pinto and Antunes 2010; Stevens and Dragićević 2007; Abolhasani et al. 2016). The land parcel is the basic morphological and functional unit in urban development, urban planning, and land policy: urban development is the change of land-use type based on the land parcel (Irwin, Bell, and Geoghegan 2003); Urban planning is a mean of spatial control based on the land parcel (Berghauser Pont et al. 2019); also formulation and implementation of land policy are based on the land parcel (Abolhasani et al. 2016). Due to the inherent morphological advantage of vector cell in characterizing cadastral plot (Jjumba and Dragićević 2012), VCAs have been able to accurately simulate land parcel's dynamic process in land-use change and have been used to provide decision support for urban planning (Barreira-González, Gómez-Delgado, and Aguilera-Benavente 2015; Abolhasani and Taleai 2020).

Aims to simulate the highly fragmented urban development processes at a realistic land parcel level, Yao et al. (2017) proposed a dynamic land parcel subdivision based vector CA (DLPS-VCA) framework. Following this framework, Zhai et al. (2020) combined vector subdivision mechanism with a CNN algorithm, and obtained a high simulation accuracy in a case study of Shenzhen, and effectively mined the relationship between multiple land-use changes and the driving factors at the neighborhood level. Vector CA based on dynamic land parcel subdivision can effectively simulate urban expansion, random fragmentation of land parcel and land-use type transition in the process of urban development, and help investigate the interaction between human activity and urban functional change with multi-source geospatial social-sensing data. Due to the complexity of the vector subdivision mechanism, DLPS-VCA has not been widely used regardless of its merits in urban simulation (Yao et al. 2017).

Besides, the lack of accuracy assessment methods specialized for VCAs inhibits the development and application of VCA modeling. Landscape index(LI) is a widely recognized and effective method to evaluate the similarity between simulation results and real landscape patterns, hence popular in accuracy assessments (Herold, Goldstein, and Clarke 2003; X. Liu et al. 2010; McGarigal, Cushman, and Ene 2012). However, considering the complexity of vector data structure in LI calculation, the existing VCA studies on land-use change simulation all conducted accuracy assessment by converting vector parcel data into raster grid data (Yao et al. 2017). Non-negligible loss of accuracy in converting vector data to raster data is thereby introduced. Such implementation would either cause the loss of accuracy or cost high computational resources if not choose the conversion scale properly (Barreira-González, Gómez-Delgado, and Aguilera-Benavente 2015). Accuracy assessment issue remains to be better addressed for calculating LI directly on vectors, which involves merging operation of vector land parcels.

Based on previous studies, an UrbanVCA framework supporting land-use change simulation at the land-parcel level is designed. This study firstly proposed and implemented a vector-based landscape index (VecLI) to allow immediate accuracy assessment for vector land parcels in tandem with UrbanVCA. In this study, Shunde, a district of Foshan City in Guangdong Province, was selected as the research area to simulate the future urban land-use change at the land-parcel scale. The effectiveness of the proposed UrbanVCA and VecLI were validated. This study also designed three scenarios to explore the changes in urban land-use patterns, including disordered development, farmland protection, and ecological protection.

## 2. Study area and data

Shunde is located in the center of Pearl River Delta Economic Zone, the richest area in Guangdong Province (Figure 1). Shunde covers an area of 806.13 km$^2$ with a permanent resident population of 2.70 million. According to the published 2016-2019 statistical yearbook, the GDP of Shunde in 2019 is 316.39 bn Yuan, which has increased by 13.27% during the period of 2016-2018. Shunde remains prosperous since the 1970s while still keeps its nature as a small-sized township city. Its complex, fragmented land-use parcels and land-use pattern, as well as its ongoing urban expansion and urban renewal projects, make it suitable for the validation of the proposed UrbanVCA framework.

Shunde owns ten township-level divisions (fourth-level administrative unit in China), whereas each subdistrict/town differs significantly in economic development modes (Wei and Zhang 2012). Besides the Daliang subdistrict, towns in Shunde are documented as specialized towns, meaning that each town has its own specialized pillar industry. Daliang subdistrict (GDP = 53.42 bn, GDP growth = +22.63%) is the downtown area of Shunde, where locates Shunde's township government. Beijiao town (60.13 bn, +16.75%), Chencun town (23.01 bn, +40.92%), and Lunjiao town (22.00 bn, +55.17%) are heavily influenced by Guangdong's capital city Guangzhou due to their close proximity and have high economic growth rates approaching or even exceeding the downtown area. Lecong town (26.11 bn, +42.65%), also namely Dongping Newcity, has set to be the future CBD area. Longjiang town (25.53 bn, +13.32%) and Leliu town (31.07 bn, +22.45%) are areas that still follow traditional specialized-township economics. Junan (17.87 bn, +21.06%) and Xintan (21.80, +4.99%) are suburbs of Shunde where have been set up high-tech industrial zones recently. Their developments are highly dependent on governmental support. Ronggui (35.43 bn, -17.22%) is a very unique area and mainly locates urban villages and traditional industries (Figure 1C). As the earliest developed area since 1970s, Ronggui's economic is almost saturated and lacks appropriate land resources for further development which even shows the trend of urban shrinkage (Y. Long and Wu 2016; Lang, Deng, and Li 2020).

Land-use changes rapidly in Shunde. From 2012 to 2018, the number of land parcels in Shunde increased by 40.49% from 16,611 to 23,336, showing a fragmentation trend in landscape. As show in Figure 1C and 1D, we classified Shunde's land-use into three types as unused land, farmland and construction land. As shown in Table 1, the land-use change in the study area mainly presents as the conversion from unused land to farmland, from unused land to construction land, and from farmland to construction land. Influenced by cultivated land requisition-compensation balance policy (Yansui Liu, Fang, and Li 2014) and urban shrinkage phenomenon in Ronggui (Fu, Yang, and Li 2020), there is a small amount of construction land transformed into farmland. The ecological protection red line area (Figure 1B) of the study area includes rivers, lakes, artificial water bodies, woodland and grassland, which accounts for 42.12% of the total area of Shunde district.

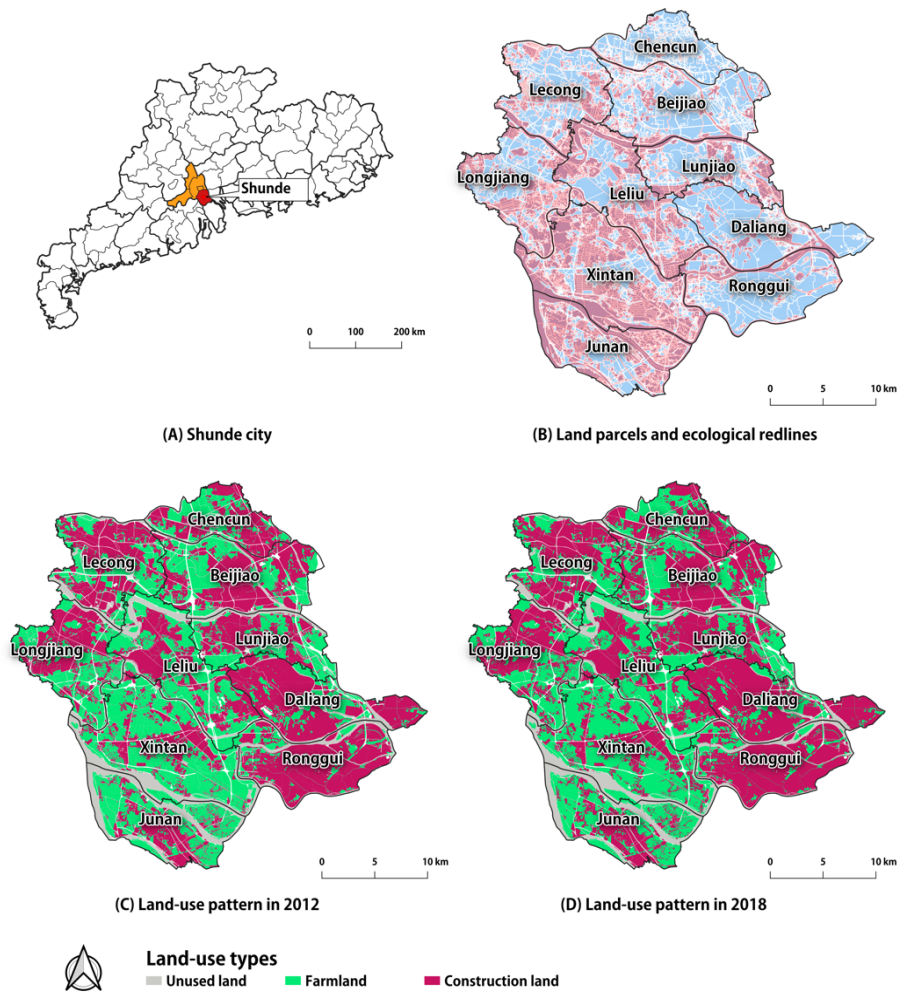

Figure 1 The case study area and its land-use patterns in 2012 and 2018.

Table 1 The land-use changes in the study area from 2012 to 2018.

| 2012 \ 2018 | Unused land | Farmland | Construction land |
|---|---|---|---|
| Unused land | 10.14% | 0.72% | 0.34% |
| Farmland | 0.12% | 37.58% | 3.44% |
| Construction land | 0.12% | 1.38% | 46.16% |

The driving factors of urban land-use change considered in this study mainly include topography, transportation facilities, industry facilities, commerce facilities and residential facilities (Verburg et al. 2002; Li and Yeh 2000; He et al. 2018; Yao et al. 2017). The data are mainly from Gaode's points-of-interest (POI) dataset, OpenStreetMap (OSM) dataset, and remote sensing imagery products. As shown in Figure 2, there are totally 17 types of spatial auxiliary data used in this study. The resolution was set to 30 meters, spatial proximity was determined by Euclidean distance or MISE-criterion-based Gaussian kernel density (Yuan, Zheng, and Xie 2012; Wand and Jones 1994). All numeric data value was normalized to range [0, 1].

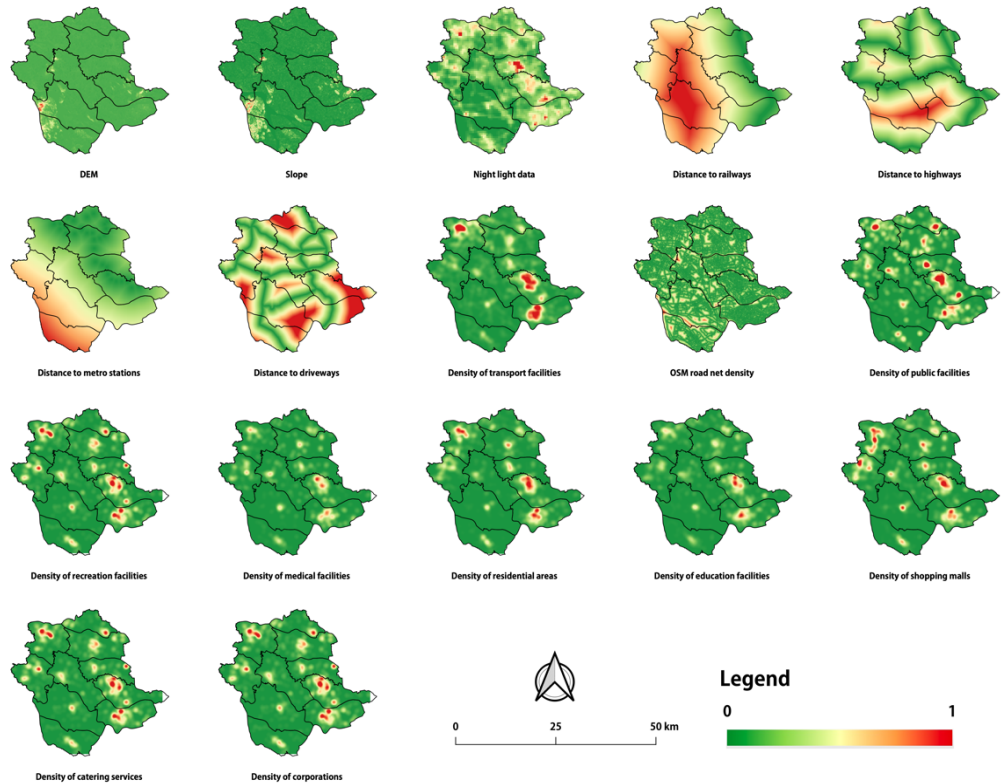

Figure 2 The spatial variables for urban development simulation in the study area.

## 3. Methodology

The proposed model consists of UrbanVCA and VecLI (Figure 3). VecLI is a landscape index calculation model based on vector land parcels, which will be described in detail in Section 3.3. The construction of UrbanVCA contains four main parts: (1) a subdivision method for obtaining the minimum vector land-parcel as basic cell unit; (2) a CA model for mining the overall development probability and simulating land-use change of vector-based cells; (3) a set of assessment methods for assessing the performance of UrbanVCA, including the proposed VecLI and conventional FoM metric; (4) an application for future land-use pattern prediction under multi-scenarios.

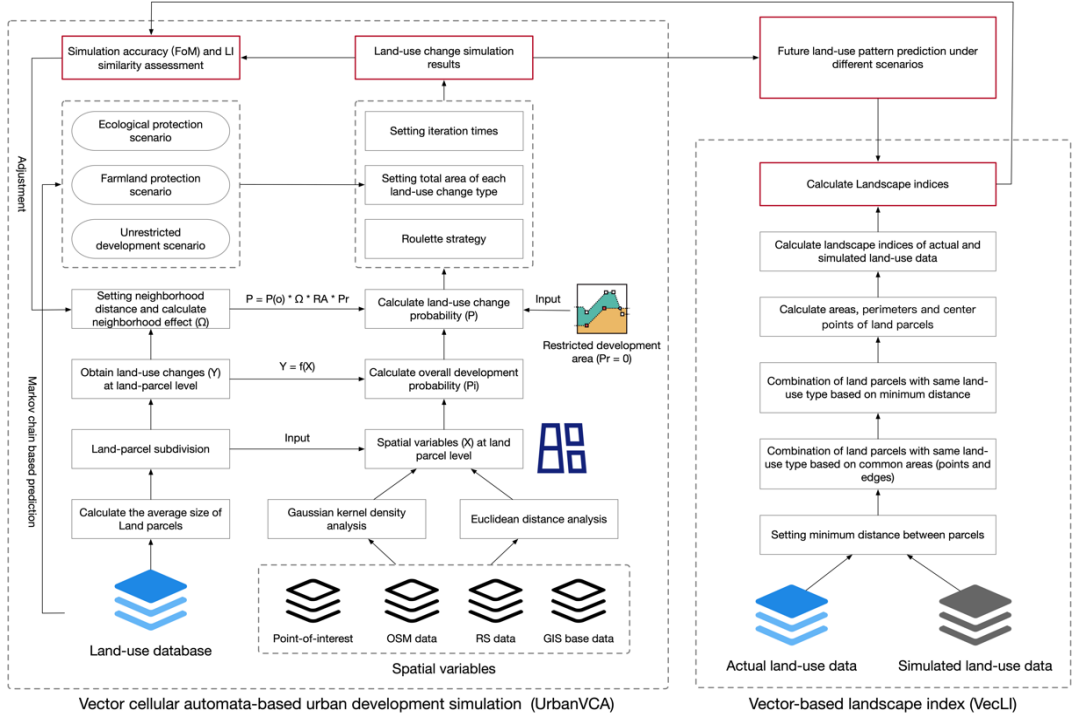

Figure 3 The workflow of the proposed UrbanVCA and VecLI in this study.

**3.1 Deriving the minimum land-parcel for UrbanVCA**

To enable a vector cell lattice, the iterative dichotomy strategy was used to split the plots (Yao et al. 2017). The iteration and splitting process will continue until the area of each plot is less than the initial state's average area of input plots and thereby get the basic cell unit. After splitting, the attribute of each cell, i.e., land-use type, was coded and then generated the minimum land-parcel data for UrbanVCA.

**3.2 Urban land-use change simulation by vector-based land parcels**

We define the spatial variable of each minimum land-parcel as X, the land-use type after land-use transformation as Y, to construct the land-use change model $Y = f(X)$. The overall development probability of land parcel $P_{(o)}$ is set as the probability that falls into each conversion type $Y_i$ in the initial year.

In this study, three machine learning algorithms, including Logistic Regression (LR), Neural Network (NN) and Random Forest (RF), were provided to calculate the overall development probability for each minimum land-parcel $P_{(o)}$. LR is a type of multivariate analysis model, which allows building multivariate regression relationship between independent and non-independent variables and conduct prediction (Yalcin et al. 2011). By adjusting the weight of each auxiliary spatial variable, LR could achieve satisfying a linear fitting result. NN is composed of layers and neurons that simulate the network of the human brain and has a good fitting result in the nonlinear mapping (Li and Yeh 2002). By continuously comparing different iteration's simulation accuracy and re-adjust the hyperparameter like the size of hidden layers and folds, NN could be thus trained and determined. RF is an aggregation of decision-tree algorithm, which uses bagging to conduct prediction or classification tasks (Mellor et al. 2013). We randomly selected 70% of the data for model training, while the rest was used for cross-validation. The optimal number of decision trees was determined by constantly adjusting the parameter and compared the corresponding simulation accuracy result.

Land-use change is affected by its neighborhood. How to determine the radius and define neighborhood effects remains a complicated problem in CA models. In this study, by taking each land-parcel as the focal unit, we used the centroid-intercepted buffer neighbourhood to obtain the optimal radius value for the determination of neighborhood effect $\Omega$ (Abolhasani et al. 2016). At time step t, neighborhood effect $\Omega$ the $j-th$ land parcel exerts on the $i-th$ land parcel could be derived mathematically as follow:

$$\Omega_{i,j}^t = e^{-d_{ij}/d} \cdot \frac{S_j / S_i}{S_{max} / S_{min}}$$

Where e is the exponential constant; $d_{ij}$ is the centroid distance between the $i-th$ land parcel and $j-th$ land parcel; $S_i$ and $S_j$ denote the area of the area of $i-th$ and $j-th$ land parcel; $S_{max}$ and $S_{min}$ represent the maximum and minimum area within the study area, respectively.

In order to determine the optimal radius value, we set the search step as 100m in the range [100, 1000] to conduct simulation. The optimal value of $\Omega$ is therefore determined by the best simulation result according to the FoM metric.

To restrict the development in water body and road, the restricted development factor Pr is set up. At time step t, the Pr value of the $i-th$ land parcel is:

$$P_{r_i}^t = \begin{cases} 0 \text{(where is restricted development areas)} \\ 1 \text{(where is not restricted development areas)} \end{cases}$$

Considering the uncertainty of land-use change process, we introduce the random factor $RA = 1 + (-\ln y)^\alpha$, where $\alpha$ is a parameter ranging within [1, 10], and $y$ denotes a random number ranging from 0 to 1. Therefore, the final land-use change probability of $i-th$ land-parcel at time step t could be derived by:

$$P_i^{k,t} = P_{(o)_i}^{k,t} \times \Omega_i^{k,t} \times P_{r_i}^t \times RA$$

Where $P_i^{k,t}$ is the total change probability developed into the $k-th$ land-use type, $P_{(o)_i}^{k,t}$ is the overall development probability developed into the $k-th$ land-use type, $\Omega_i^{k,t}$ is the neighborhood effect, $P_{r_i}^t$ is the restriction development factor, and RA is the random factor.

In the simulation process, we set the increment number of cells in an iteration as N / M based on the land-use increment N and iteration number M of the target year. The dead cycle situation in the iteration process is prevented through the control of the development probability threshold. Due to the complex, chaotic system nature of urban development, we introduce the Roulette strategy to allow the chance that land-parcel could turn to a land-use type that has a lower development probability. (Liu et al. 2017)。

Specifically, in this study, neighbourhood distance was set 600 meters, data splitting ratio of training data and testing data was set to be 4:1. In the NN-based model, the number of input nodes (17) is consistent with the number of spatial variables (as shown in Figure 2), the number of output nodes (3) is consistent with the number of conversion types (unused land, farmland, and construction land). The number of the hidden layer is set to 3, and 10-fold cross validation is used to obtain the optimal urban land-use change simulation model. In the RF-based model, the number of decision trees is set to 80 (Zhang et al. 2020; Yao et al. 2019; Oshiro, Perez, and Baranauskas 2012). The best fit model was obtained by OOB estimation with an OOB rate setting as 0.3.

### 3.3 Simulation accuracy assessment using VecLI

VecLI is a calculation and analysis model of landscape index specialized for vector-based land parcel, which can be used to calculate Figure-of-Merit (FoM) and LI similarity. Considering the possible topological errors occurred in the land-parcel subdivision process, VecLI is able to quickly conduct topology check and merge the adjacent parcels with similar land-use type to prevent needless subdivision and excessive fragmentation of land-parcels.

Due to the topological complexity introduced by scattered, fragmented, and tiny parcels in LI computation, two operations were conducted to mitigate this problem: first, a smaller neighborhood radius was set; second, the parcels whose area was less than three times the standard deviation was filtered. After topology checking of vector data, based on the area and perimeter of the merged parcel, and the centroid coordinate of the parcel before merging, this study selected the number of parcels (NP), largest-parcel index (LPI), mean Euclidean nearest-neighbor distance (ENN) and the mean perimeter–area ratio (PARA) to evaluate the similarity between simulation results and the ground-truth land-use patterns.

FoM metric was used for accuracy assessment in this study. The formula is as follows:
$$FoM = B / (A + B + C + D)$$
$$Product's\ accuracy\ (PA) = B / (A + B + C)$$
$$User's\ accuracy\ (UA) = B / (B + C + D)$$

Where A denotes parcel that remains unchanged in simulation while in ground-truth the parcel have changed; B denotes parcel that correctly predicts land-use change as well as the land-use type; C denotes parcel that correctly predicts land-use change however with a wrong land-use type; D represents parcel that has land-use change in simulation while in ground-truth the parcel remains unchanged.

LI similarity was used to measure the land-use pattern similarity between the actual and simulated land-use in this study. The formula is as follow:
$$\alpha_l = 1 - \frac{1}{n}\sum_i \Delta l_i$$
$$\Delta l_i = \begin{cases} |l_{i,s} - l_{i,o}|/l_{i,o}, & l = NP, ENN, PARA \\ |l_{i,s} - l_{i,o}|, & l = LPI \end{cases}$$

Where $l_{i,s}$ and $l_{i,o}$ represent the $i-th$ LI of the simulated and real scenarios, respectively. $\Delta l_i$ is the normalized difference of the $i-th$ LI. $\alpha_l$ is the landscape pattern similarity between the actual and simulated land-use, and n refers to the number of LIs.

### 3.4 Urban land-use change prediction under multi-scenarios

The Markov chain model was used to predict the future area of each type of land-use (Sang et al. 2011). Three development scenarios were designed in this study: (1) Unrestricted development scenario (S1); (2) Farmland protection scenario (S2); (3) Eco-protection scenario (S3).

In scenario S1, both unused land and farmland were allowed to transform into construction land; in scenario S2, it was forbidden to turn farmland into construction land; in scenario S3, all farmlands and lands within the ecological red line were prohibited from developing into construction land. Constrained Markov Chain was used in both the S2 and S3 protection scenarios to derive the total area of land-use transformation.

Land-use change for 2030 is simulated based on the above three scenarios, and both qualitative and quantitative research were conducted by exploring the spatial pattern of land-use and the change of landscape index.

## 4. Results

4.1 Simulation results and accuracy assessment

In this study, the land-use changes from 2012 to 2018 in the study area were simulated, and the land-use change in 2015 was used to validate the result. Table 2 and Table 3 show the simulation accuracy and landscape index similarities of the proposed UrbanVCA model, respectively. RF-based model achieves the highest simulation accuracy, which is 69.93% and 228.38% higher than NN-based and LR-based models, respectively. It shows that there is a complex nonlinear correlation among spatial variables, which leads to the poor accuracy of logistic regression. In addition, land-use transformation rules differ greatly for different regions of the study area, where RF can better mine a variety of transformation rules than NN (Yao et al. 2017; Zhai et al. 2020).

Table 2 The simulation accuracy (FoM) of different machine learning methods based on UrbanVCA models.

| Years | Accuracy | LR-based | NN-based | RF-Based |
|---|---|---|---|---|
| 2018 | FOM | 0.074 | 0.143 | 0.243 |
| 2018 | UA | 0.173 | 0.270 | 0.433 |
| 2018 | PA | 0.114 | 0.230 | 0.355 |
| 2015 (Validation) | FOM | 0.049 | 0.073 | 0.112 |
| 2015 (Validation) | UA | 0.173 | 0.217 | 0.274 |
| 2015 (Validation) | PA | 0.064 | 0.098 | 0.158 |

From Table 3, it can be found that the RF-based model still gets satisfying results as both the LPI and ENN metric are very close to the actual case. Interestingly, LI similarities obtained in 2015 as validation are generally higher than the simulation result of 2018. According to the official statistical yearbook, the GDP growth rate of Shunde from 2015 to 2018 (22.63%) is 2.19 times than the period 2012-2015 (10.31%). We assumed that Shunde's fast growth of GDP has led to dramatic changes in land parcels, both in morphological characteristics and absolute amounts. There is a certain correlation among the simulation accuracy, GDP growth, and government policies, which we will discuss in detail in Section 4.2. Due to the driving factors of government policies, the similarity of simulated landscape index in 2018 is slightly lower than that in 2015, and we assume that the government policy influenced the actual process of urban development greatly, which generated a much stronger fragmentation degree of land than that of our simulation results, and a higher perimeter/area ratio (PARA) than that of the simulation results.

Table 3 The landscape indices and similarities between the actual land-use and simulated land-use patterns via UrbanVCA.

| Years | Models | NP | LPI | ENN | PARA | LI Similarity |
|---|---|---|---|---|---|---|
| 2018 | Actual | 13502 | 0.419 | 52.532 | 0.112 | / |
|  | LR-based | 10874 | 0.418 | 45.111 | 0.086 | 0.859 |
|  | NN-based | 11161 | 0.426 | 47.081 | 0.085 | 0.869 |
|  | RF-based | 11023 | 0.418 | 48.029 | 0.085 | 0.873 |
| 2015 (Validation) | Actual | 12432 | 0.430 | 52.965 | 0.103 | / |
|  | LR-based | 10874 | 0.418 | 45.111 | 0.086 | 0.894 |
|  | NN-based | 11159 | 0.424 | 47.197 | 0.085 | 0.905 |
|  | RF-based | 11038 | 0.423 | 47.850 | 0.085 | 0.906 |

4.2 The relationship between economic level and simulation accuracy

Figure 4 shows the spatial distribution of simulation errors for each town/subdistrict in the study area. The simulation error of the RF-based model and NN-based model is smaller than that of the LR-based model. Combined with Figure 5D, it could be found that compared with the suburbs (Junan and Xintan), the periphery belt area around Guangzhou (Lunjiao, Beijiao, and Chencun) and the Dongping Newcity (Lecong) has shown extremely high simulation accuracy with an average of FoM exceeding 0.4(Pontius et al. 2008). The development of land-use in Ronggui has reached its limit, and its construction land has even degenerated into unused land. Urban land degradation is an extremely rare phenomenon in the context of China's rapid development environment(Yansui Liu and Wang 2019). UrbanVCA did not consider the degradation of land-use, which leads to low simulation accuracy in Ronggui (FoM = 0.070).

Through the comparative analysis of FoM-GDP correlation and FoM-GDP growth rate correlation from Figure 5, it is clear that the simulation accuracy of UrbanVCA has little to do with the area's economic status, while the area's rapid growth strongly correlates with simulation accuracy. The linear fitting result shows that the correlation between FoM and GDP is only 0.071 (Figure 5C), while the correlation between FoM and GDP growth rate is 0.480 (Figure 5B). The development of Xintan and Junan is mainly driven by government policies. For example, there is an "enclave" in the southwest corner of Xintan, which is a high-tech industrial zone planned by the government. The "Top-Down" governmental planning greatly influences the development process, which leads to low simulation accuracy. If exclude Xintan, Junan, the correlation between simulation accuracy and GDP growth rate can reach as much as 0.645 (Figure 5B). It shows that UrbanVCA can effectively simulate the urban development process and positively correlate with the economic development trend with high fitting accuracy while keeping the vector-based land parcel for a more realistic morphological representation of geographic entities.

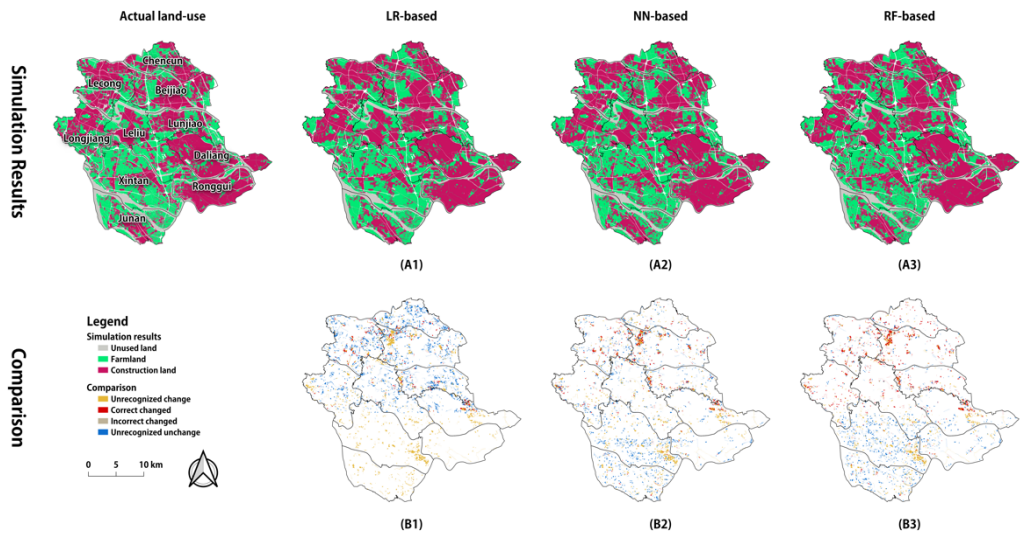

Figure 4 The different machine learning based simulation results (A1~A3) and the comparison between actual urban land-use and simulation results (B1~B3).

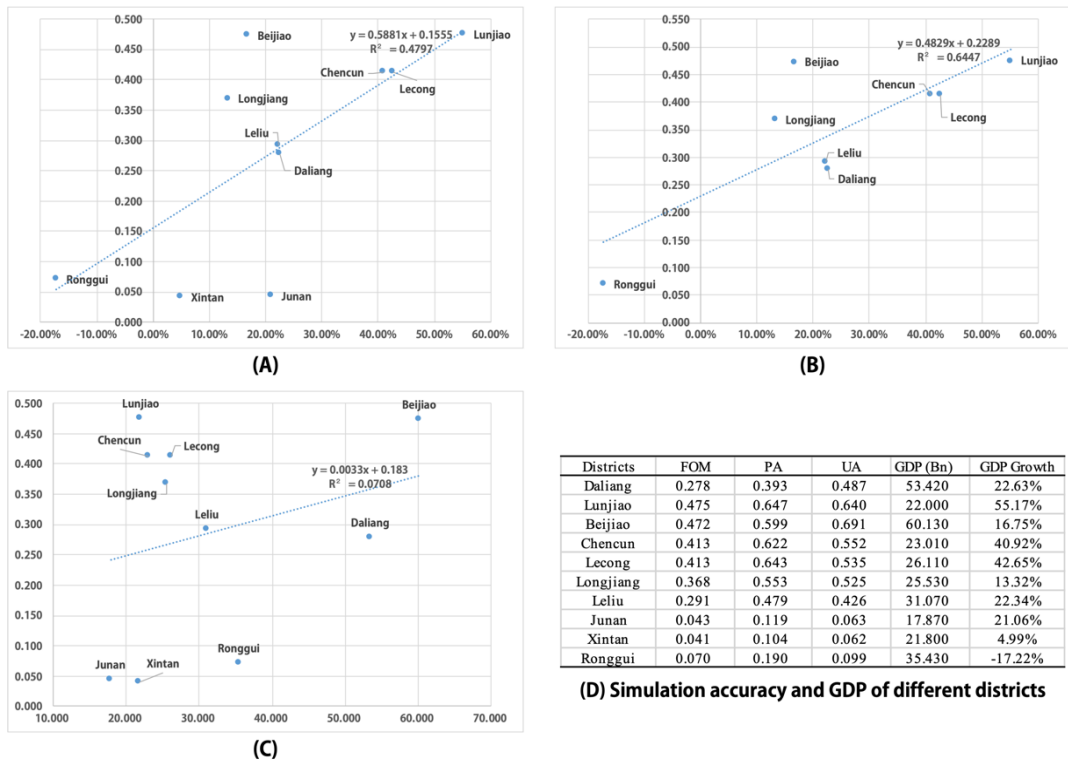

| Districts | FOM | PA | UA | GDP (Bn) | GDP Growth |
|---|---|---|---|---|---|
| Daliang | 0.278 | 0.393 | 0.487 | 53.420 | 22.63% |
| Lunjiao | 0.475 | 0.647 | 0.640 | 22.000 | 55.17% |
| Beijiao | 0.472 | 0.599 | 0.691 | 60.130 | 16.75% |
| Chencun | 0.413 | 0.622 | 0.552 | 23.010 | 40.92% |
| Lecong | 0.413 | 0.643 | 0.535 | 26.110 | 42.65% |
| Longjiang | 0.368 | 0.553 | 0.525 | 25.530 | 13.32% |
| Leliu | 0.291 | 0.479 | 0.426 | 31.070 | 22.34% |
| Junan | 0.043 | 0.119 | 0.063 | 17.870 | 21.06% |
| Xintan | 0.041 | 0.104 | 0.062 | 21.800 | 4.99% |
| Ronggui | 0.070 | 0.190 | 0.099 | 35.430 | -17.22% |

(D) Simulation accuracy and GDP of different districts

Figure 5 The correlation between FoM (y-axis) and GDP growth rates (A), GDP growth rates (B, without policy-led new technology development zones) and GDP (C).

4.3 The influence of neighbourhood effect on UrbanVCA model accuracy

Compared with the traditional raster-based CA, the neighbourhood effect of VCA is more complex and important (Dahal and Chow 2015; Abolhasani and Taleai 2020). UrbanVCA adopted centroid-intercepted buffer neighbourhoods to compute the neighbourhood influence probability of each land parcel. Figure 6 shows that the accuracy of VCA based simulation results is very sensitive to the setting of neighbourhood distance. When the neighbourhood setting is small (less than 200 metres), the accuracy of FoM keeps an improving trend. When the distance is set to 400-600 metres, FoM is steady above the value of 0.2 with a PA value greater than 0.4. With the increase of neighbourhood setting, the simulation accuracy shows an oscillation curve and declines. Therefore, the neighbourhood distance of 600 metres was set in this study. It is worth noting that previous studies confirmed that the internal structure of urban functional areas generally presents an spatial aggregation pattern (Tu et al. 2017; Shen and Karimi 2016; Yao et al. 2017), which means that a region's functional change will be impacted by the adjacent urban structure within a certain distance range, while beyond this range, such impact will weaken measurably with the increase of distance.

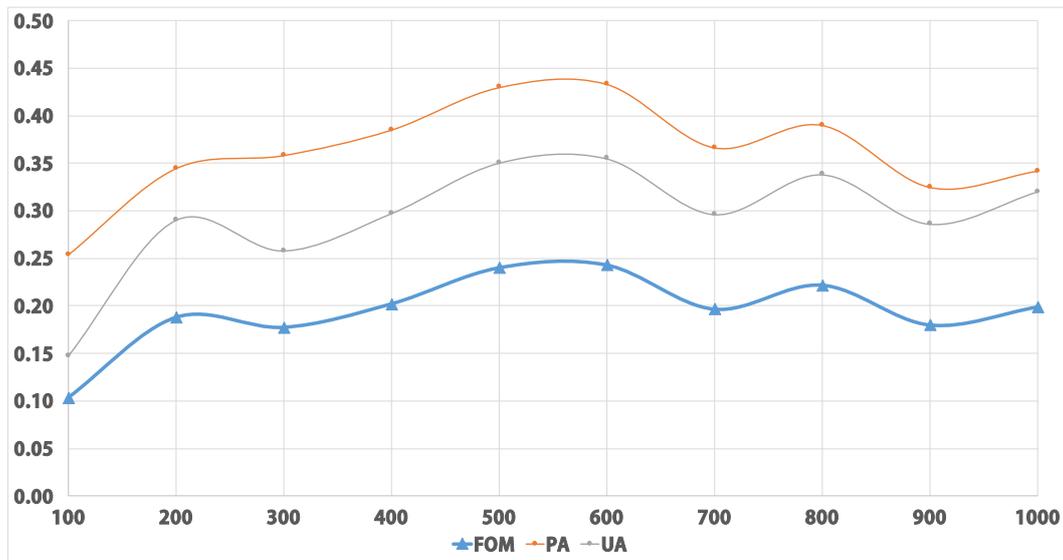

Figure 6 The relationship between the neighborhood buffer distance (x-axis, unit: meter) and simulation accuracy (y-axis).

4.4 Simulation results in 2030 under different scenarios

Table 4 shows the total amount of various types of land-use in 2013 predicted by the Constrained Markov Chain. The simulation results based on UbranVCA are shown in Figure 7, where the blue parcels are lands transformed to construction land from unused land or farmland. Table 5 shows the landscape index of the simulated land-use patterns under three different scenarios. Combined with Figure 7 and table 5, it can be found that compared with disordered development mode, the fragmentation degree of farmland or ecological protection scenarios has decreased by 1.95% and 2.48%, and LPI has increased by about 1.00% ~ 2.25%. In other words, under the protection scenarios (S2 and S3), urban development is mainly concentrated around the existing construction land where "enclave" development hardly happens, and agglomeration trend is strengthened, which is conducive to reducing the allocation cost of public and commercial resources in the city (Fang et al. 2018; H. Long et al. 2016; Shu and Xiong 2019).

Table 4 The total area (unit: km$^2$) of each urban land-use in 2030 under different scenarios.

| Scenarios | Unused land | Farmland | Construction land |
|---|---|---|---|
| S1 | 83.835 | 257.147 | 465.152 |
| S2 | 82.222 | 293.806 | 430.106 |
| S3 | 87.553 | 291.884 | 426.696 |

Table 5 The landscape indices of land-use pattern simulation results in 2030 under different scenarios.

| Scenarios | NP | LPI | ENN | PARA |
|---|---|---|---|---|
| S1 | 11562 | 0.399 | 43.209 | 0.079 |
| S2 | 11337 | 0.403 | 45.467 | 0.081 |
| S3 | 11275 | 0.408 | 45.589 | 0.081 |

From the perspective of farmland and ecology protection (Figure 6), the current development mode in the study area will lead to a large number of losses in artificial water bodies, forests, and grassland in 2023. The ecological protection area and farmland occupation will reach 14.36% and 10.58%, respectively. Under the scenario of farmland protection, 57.95% of ecological environment damage and 66.45% of farmland occupation can be reduced. Under the scenario of ecological protection, the ecological environment damage could be reduced by 62.04% compared with the current development mode. In S2 and S3 scenarios, damage and occupation of the ecological zone and farmland mainly occurred around the existing construction land. The erosion of the natural water body will be about 1% by 2030 in all three scenarios. Therefore, for sustainable development, farmland or ecological protection should be considered in urban planning to protect basic farmlands and promote a compact development mode for the city,

Table 6 Erosion rates of economic protection zones and farmland in 2030 under different scenarios.

| Scenarios | Natural waterbody | Artificial waterbody | Forest land | Grassland | Eco-zone (Total) | Farmland |
|---|---|---|---|---|---|---|
| S1 | 1.10% | 18.67% | 16.74% | 27.85% | 14.36% | 10.58% |
| S2 | 1.10% | 7.24% | 4.08% | 45.00% | 6.04% | 3.55% |
| S3 | 0.95% | 7.09% | 3.50% | 13.58% | 5.45% | 3.54% |

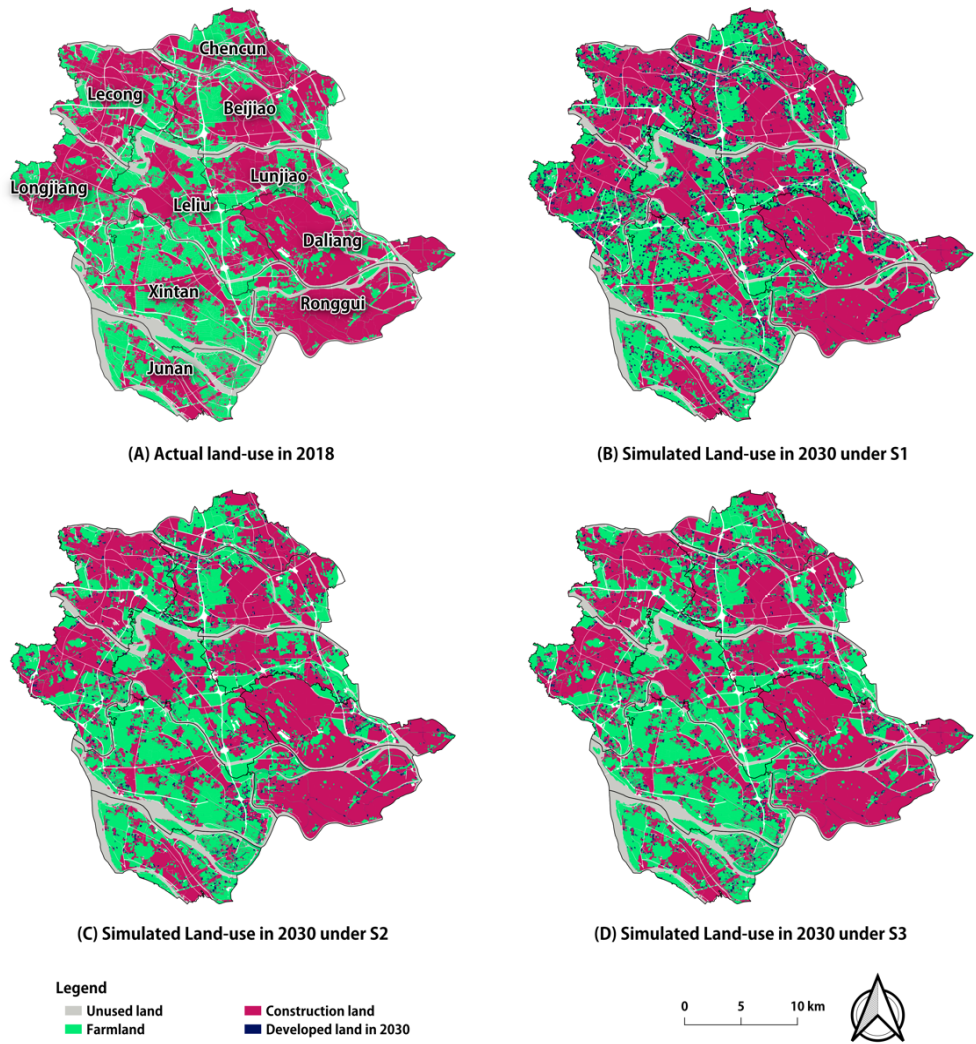

Figure 7 The simulation results in 2030 under different scenarios, including (B) Unrestricted development scenario (S1), (C) Farmland protection scenario (S2) and (D) Ecosystem protection scenario (S3).

## 5. *Discussion*

Aiming at simulating a land-use change in urban renewal scenario and addressing the challenge in apply DLSP-VCA, we proposed UrbanVCA, an easy-to-use framework, at a realistic land parcel level for the first time. UrbanVCA can not only simulate the process of land fragmentation but also support a variety of machine learning algorithms to mine the probability of urban land-use change. At present, there is no method that directly calculates the LI on vector-based parcel data. The VecLI was proposed in order to avoid the accuracy loss aroused by conventional rasterization landscape index calculation. By merging adjacent parcels with similar land-use types in the vector space, VecLI is able to ensure the calculating accuracy and directly derive the LI at a vector cell level, and obtain the most reasonable LI for land-use pattern similarity analysis.

The proposed UrbanVCA framework was used to simulate the urban land-use change in Shunde,

Foshan City. Previous studies have pointed out that land-use change within the city is based on the irregular shaped land parcel where raster-based CA could not well apply (Zhai et al. 2020; Yao et al. 2017). Therefore, we did not carry out a comparison experiment between raster-based CA and the proposed UrbanVA. Three popular machine learning models are intergrated into UrbanVCA. The simulation results show that the RF-based model has the best simulation accuracy (FoM = 0.243) in the study area. The LR-based model performed poorly because spatial variables have complex nonlinear correlation (Fouedjio and Klump 2019). Considering the strong interpretability merit of LR-based model, the selection of spatial variables could be carefully designed and leverage such advantages in future research.

This study also found that there is a strong correlation between the accuracy of urban land-use simulation and GDP growth rate. In the case that strong policy intervention is missed, UrbanVCA has been able to deliver accurate simulation results (FoM > 0.25). In fast-growing areas, the FoM could even exceed 0.4, which is already surpassed the results of state of the art research (Pontius et al. 2008; Zhai et al. 2020). In the meantime, FoM and GDP growth rate in the study area have shown a significant positive correlation ($R^2$=0.645), whereas the simulation accuracy decreased sharply in scenarios with strong policy intervention (such as the newly established high-tech economic zone). How to effectively quantify and integrate policy factors into transition rules mining in the CA model has always been a challenge in urban simulation research (D. Chen et al. 2019) and would be further studied in future research of UrbanVCA.

The sensitivity analysis of neighborhood distance settings has shown an interesting phenomenon. With the increase of neighborhood distance setting, the simulation accuracy first improves and then decreases, which indicates the existence of function aggregation within the city and is consistent with previous VCA studies (Zhai et al. 2020; Yao et al. 2017). Previous studies have found that Shenzhen (one of the developed mega-city in southern China) has the best neighborhood setting at 800 meters. In this study, we found that the best neighborhood setting for Shunde is 600 meters, which indicates that small cities have a stronger aggregated pattern of urban function compared with mega-cities, and area with complete urban functions tends to be smaller in small cities. Such findings provide a new way to analyze the average size of urban functional areas, which requires further discussion in the future.

This study also carried out a land-use change simulation for 2030 and compared the impact of urban development on ecological protection area and farmland erosion under three different scenarios. Under the scenario of farmland or ecological protection, internal urban development presents a stronger agglomeration effect, which is conducive to reducing the cost of public and commercial resource allocation and planning (Tu et al. 2017). Meanwhile, under the protection scenario, the erosion of farmland and the ecological environment in the study area could be reduced by about 60%. This result supports the necessity to take ecological and farmland protection instead of disordered, short-term development.

There are several notable limitations of this study. First, the degradation phenomenon that occurred in construction land did not get careful consideration. At present, either previous CA models or the proposed UrbanVCA would attach importance to the urbanization of non-construction land, whereas very limited concern was shown on the urban shrinkage issue and model the transition from construction land to other types of land-use. Second, there is a strong correlation between urban land-use change and human activities(Lin, Wu, and Li 2019; Seto et al. 2012). Urban land-use simulation without social and human activity modelling would ignore one of the most important driving factors in the land-use change process. Therefore, it is an important research direction to investigate urban parcels' telecommunication effect and quantitatively integrate it into the UrbanVCA framework in the future.

## *6. Conclusion*

To solve the problem that the vector-based CA model is difficult to implement, this study proposed an UrbanVCA framework for the first time and developed an effective VecLI system to calculate and evaluate landscape indices using vector-based data. It is found that without strong policy intervention, RF-based UrbanVCA can effectively mine the transition rules of urban land-use change. Meanwhile, results indicate a significant positive correlation between the simulation accuracy of urban land-use change and the speed of regional economic development. By simulating the future land-use change scenarios, it can be found that ecological and farmland protection policies can effectively optimize the natural environments and human settlements simultaneously. On the basis of this study, we developed and released UrbanVCA software to provide urban planners and researchers an effective tool to conduct urban development research. Supplementary materials, including user manual and video tutorial, could be referenced for detailed usage and design principle information. In future research of UrbanVCA, we will work on policy factors quantifying, urban shrinkage modeling, and telecommunication effect modeling at the actual land parcel level to improve the simulation accuracy and versatility of UrbanVCA.